\documentclass[prl,superscriptaddress,showpacs,longbibliography,reprint]{revtex4-2}

\usepackage[colorlinks=true,linkcolor=blue,urlcolor=blue,citecolor=blue,pdfusetitle]{hyperref}
\usepackage{color}
\usepackage[utf8]{inputenc}
\usepackage[usenames,dvipsnames]{xcolor}
\usepackage{amsmath}
\usepackage{amssymb}
\usepackage{graphicx}
\graphicspath{{graphics/}}
\usepackage{epsfig}
\usepackage{dcolumn}
\usepackage{bm}
\usepackage{mathrsfs}
\usepackage{multirow}
\usepackage[all]{xy}
\usepackage{pbox}
\usepackage{lipsum}
\usepackage{verbatim}
\usepackage{braket}
\usepackage{dsfont}
\usepackage{array}
\usepackage{makecell}
\usepackage{tabularx}
\usepackage{isotope}
\usepackage{xr}
\usepackage[normalem]{ulem}

\usepackage[colorlinks=true,linkcolor=blue,urlcolor=blue,citecolor=blue,pdfusetitle]{hyperref}

\begin{document}

\title{Quantum trajectories of dissipative time-crystals}
\author{Albert Cabot}
\affiliation{Institut f\"ur Theoretische Physik, Eberhard Karls Universit\"at T\"ubingen, Auf der
Morgenstelle 14, 72076 T\"ubingen, Germany.}
\author{Leah Sophie Muhle}
\author{Federico Carollo}
\affiliation{Institut f\"ur Theoretische Physik, Eberhard Karls Universit\"at T\"ubingen, Auf der
Morgenstelle 14, 72076 T\"ubingen, Germany.}
\author{Igor Lesanovsky}
\affiliation{Institut f\"ur Theoretische Physik, Eberhard Karls Universit\"at T\"ubingen, Auf der
Morgenstelle 14, 72076 T\"ubingen, Germany.}
\affiliation{School of Physics, Astronomy, University of Nottingham, Nottingham, NG7 2RD, UK.}
\affiliation{Centre for the Mathematics, Theoretical Physics of Quantum Non-Equilibrium Systems,
University of Nottingham, Nottingham, NG7 2RD, UK}
\begin{abstract}
Recent experiments with dense laser-driven atomic gases [G. Ferioli \textit{et al.}, arXiv:2207.10361 (2022)] have realized a many-body system which in the thermodynamic limit yields a so-called boundary time-crystal. This state of matter is stabilized by the competition between coherent driving and collective dissipation. The aforementioned experiment in principle allows to gain \textit{in situ} information on the nonequilibrium dynamics of the system by observing the state of the output light field. We show that the photon count signal as well as the homodyne current allow to identify and characterize critical behavior at the time-crystal phase transition. At the transition point the dynamics of the emission signals feature slow drifts, which are interspersed with sudden strong fluctuations. The average time between these fluctuation events shows a power-law scaling with system size, and the origin of this peculiar dynamics can be explained through a simple non-linear phase model. We furthermore show that the time-integrated homodyne current can serve as a useful dynamical order parameter. From this perspective the time-crystal can be viewed as a state of matter in which different oscillation patterns coexist.
\end{abstract}

\maketitle
The interplay between driving and dissipation can stabilize genuine non-equilibrium phases of interacting quantum systems \cite{Diehl2008,Diehl2010,Lee2013,Jin2013,Marcuzzi2014}. A manifestation, currently receiving significant attention, are time-crystals \cite{Else2020} which are many-body phases that break time-translation symmetry. These crystalline structures in time can emerge as a consequence of different mechanisms but they all manifest in stable oscillatory asymptotic regimes. This means that at long times observables display a periodicity in time that breaks either the discrete (i.e., by displaying a subharmonic response) \cite{Gong2018,Wang2018,Gambetta2019,Lazarides2020,Riera2020,Zhu2019,Chinzei2020,Kessler2021,Tuquero2022,Sarkar2022} or the continuous \cite{Iemini2018,Tucker2018,Buonaiuto2021,Lledo2020,Carollo2022,Seibold2020,Buca2019} time-translation symmetry of the dynamical generator. A setting in which such non-equilibrium dynamics can be studied is constituted by atomic ensembles interfaced with optical cavities \cite{Ritsch2013} or photonic structures \cite{Chang2018}. In these systems spatial self-organization and steady state supperradiance \cite{Domokos2002,Meiser2009,Bohnet2012,Norcia2016}, synchronization \cite{Xu2013,Zhu2015}, as well as time-crystal oscillations \cite{Tucker2018,Buonaiuto2021,Zhu2019,Kessler2020,Kessler2021,Tuquero2022} have been reported. Even in free space, i.e. without enhancing collective effects through a cavity, long-time oscillatory dynamics can be supported within dense atomic ensembles. This has been discussed in early theoretical studies on cooperative resonance fluorescence --- see e.g. Refs. \cite{Agarwal1977,Narducci1978,Drummond1978,Carmichael1980} --- long before the concept of time-crystals was established. In these dense gases coherently driven (two-level) atoms are subject to collective dissipation \cite{Dicke1954}, which ultimately stabilize time-periodic solutions. This conceptually simple system, which in the current terminology would be referred to as a (dissipative) boundary time-crystal \cite{Iemini2018,Buonaiuto2021,Carollo2022}, was recently realized in an experiment \cite{Ferioli2022}. This achievement opens up a new perspective for exploring and characterizing the time-crystal transition in a dissipative setting. This is owed to the fact that an open system allows to extract \textit{in situ} information on the quantum many-body dynamics.  

\begin{figure*}[t]
\includegraphics[width=1\linewidth]{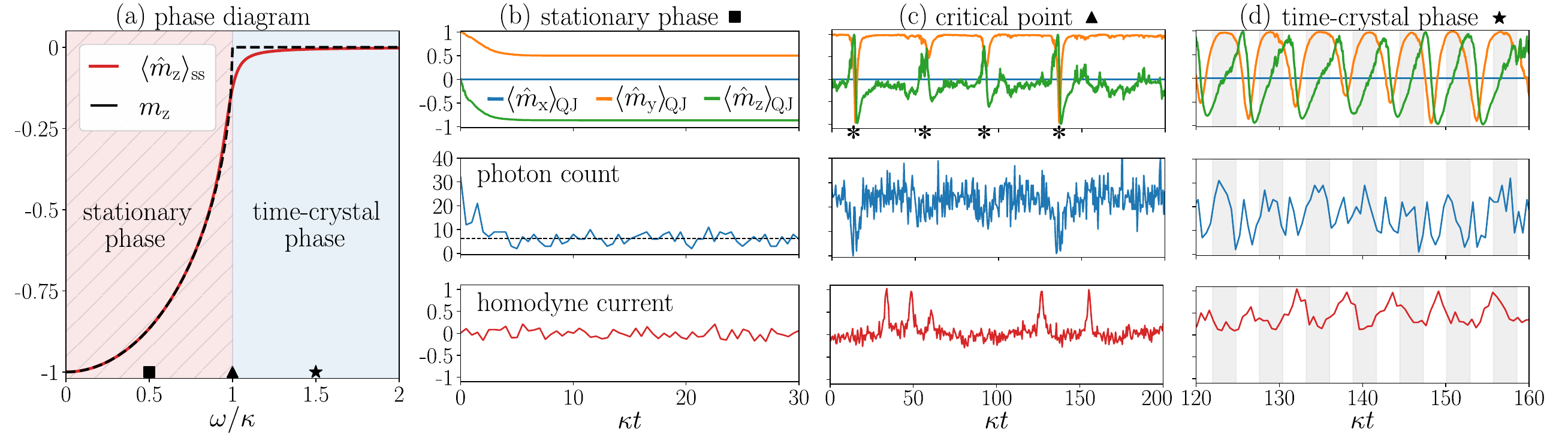}\\
\caption{\textbf{Signatures of the dynamical regimes in individual realizations.} (a) Stationary value of the $z$-component of the magnetization, $m_z$, for $N=100$ (red solid line) and time-averaged mean-field solution for $\mathcal{M}=0$ (black dashed line). At $\omega/\kappa=0.5$ ($\blacksquare$) and $\omega/\kappa=1.5$ ($\bigstar$) the system is in the stationary phase and the time-crystal phase, respectively. At the criticial point $\omega/\kappa=1$ ($\blacktriangle$) the phase transition takes place. (b-d) Representative quantum trajectories for the magnetization components (top), the photon count (middle) and the homodyne current (bottom). The photon count trajectories are computed with a bin size $\kappa\Delta t=0.5$, while the homodyne current is averaged over a sliding time windows of $\kappa\Delta t=0.5$ and rescaled by $1/\sqrt{2N}$. In all cases $N=100$ and the initial condition corresponds to a spin coherent state as defined by  $|\theta,\phi\rangle=\exp[i\theta( \hat{J}_x\sin\phi-\hat{J}_y\cos\phi)]|J,J\rangle$ ($\theta \in [0,\pi]$,  $\phi\in[0,2\pi]$) \cite{Ma2011} with $\theta=\pi/2$ and $\phi=\pi/2$.}\label{fig_regimes}
\end{figure*}

In this work, we show how the transition and the concomitant critical behaviour manifest at the level of quantum trajectories which result from continuously monitoring. We consider two different measurement schemes which yield qualitatively distinct types of quantum trajectories: the record of emitted photons and the quadrature (photo-current) of the output light-field. Both, the quantum jump trajectories of photon counting as well as the homodyne (diffusive) trajectories of the photo-current display clear signatures of the time-crystal phase transition. We show how these experimentally accessible quantities allow to identify the critical point. Here the trajectories show a peculiar dynamical behavior with small drifts being interspersed with large jumps, and the average time between jumps follows a characteristic power-law. Furthermore, we demonstrate that a convenient way for characterizing the time-crystal phase transition is through dynamical order parameters, such as the time-integrated photo-current. When coupling to the appropriate quadrature, the photo-current shows highly intermittent behavior in the time-crystal phase. This peculiarity can be interpreted as dynamical coexistence between various time-crystalline, i.e. periodic, solutions. 

\noindent {\it Time-crystal phase transition in the thermodynamic limit.--} The model we consider here consists of $N$ two-level atoms with coherent resonant driving and collective dissipation. Its dynamics is described by a Markovian master equation governing the time evolution of the density matrix $\hat{\rho}$ as ($\hbar=1$)  
\begin{equation}\label{normal_ME}
\partial_t \hat{\rho}=\mathcal{L}\hat{\rho}=-i\omega[ \hat{J}_\mathrm{x},\hat{\rho}]+\frac{\kappa}{N}\big(2\hat{J}_-\hat{\rho}\hat{J}_+-\hat{J}_+\hat{J}_-\hat{\rho}-\hat{\rho}\hat{J}_+\hat{J}_-\big).
\end{equation}
Here $\omega$ is the Rabi frequency, which parametrizes the coupling strength between the atoms and the laser and $2\kappa/N$ the collective atomic decay rate. Note, that the latter scales with $1/N$, which is necessary for obtaining a well-defined thermodynamic limit. The master equation solely depends on collective spin operators defined as $\hat{J}_\mathrm{\alpha}=\frac{1}{2}\sum_{j=1}^N \hat{\sigma}_\mathrm{\alpha}^{(j)}$ ($\alpha=\mathrm{x,y,z}$), with $\hat{\sigma}_\mathrm{\alpha}^{(j)}$ being the Pauli matrices and $\hat{J}_\pm=\hat{J}_\mathrm{x}\pm i\hat{J}_\mathrm{y}$. As shown in \cite{Carmichael1980,Benatti2018,Iemini2018,Carollo2022} it is convenient to introduce the magnetization vector with components  $\hat{m}_\alpha=\hat{J}_\alpha/(N/2)$. In the thermodynamic limit ($N\rightarrow\infty$) the dynamics of the expectation values $m_\alpha=\langle\hat{m}_\alpha\rangle$ is exactly governed by the mean-field equations $\dot{m}_\mathrm{x}= \kappa \,m_\mathrm{x} m_\mathrm{z}$, $\dot{m}_\mathrm{y}= -\omega \,m_\mathrm{z}+\kappa\, m_\mathrm{y} m_\mathrm{z}$, and $\dot{m}_\mathrm{z}=\omega\, m_\mathrm{y}-\kappa\,(m_\mathrm{x}^2+m_\mathrm{y}^2)$. The ensuing dynamics features two conserved quantities, which are the total angular momentum $j^2=m_{\mathrm{x}}^2+m_{\mathrm{y}}^2+m_{\mathrm{z}}^2$ and the quantity  $\mathcal{M}=m_\mathrm{x}/(m_\mathrm{y}-\omega/\kappa)$ (we will focus throughout on the case $j^2=1$, $\mathcal{M}=0$).

The model undergoes a non-equilibrium phase transition at the critical ratio $\omega/\kappa=1$. Below this critical point, the system approaches a unique stable stationary fixed point. Above the critical point, $\omega/\kappa>1$, it displays a continuous family of non-isolated closed orbits covering the whole Bloch sphere \cite{Drummond1978,Carmichael1980,Iemini2018}. Each orbit is associated with a value of the conserved quantity $\mathcal{M}$. The emergence of this non-equilibrium transition for finite particle number $N$ is signaled by a sharp crossover of the stationary value of the $z$-component of the magnetization, $\langle \hat{m}_z \rangle_\mathrm{ss}$, at $\omega/\kappa=1$ [see Fig.~\ref{fig_regimes}(a)].

\noindent {\it Time-crystal phase transition in finite systems.-- } The master equation \eqref{normal_ME} describes the average evolution of an open system in terms of its density matrix $\hat{\rho}$. In an experiment, however, one observes stochastic realizations of the open systems dynamics --- the so-called quantum trajectories. Averaging over these individual realizations yields the density matrix. To be specific, let us consider the dynamics of the magnetization components in the stationary phase, which are shown in the top panel of Fig.~\ref{fig_regimes}(b). These curves are obtained by a so-called quantum jump unravelling \cite{Wiseman2009} (see also \cite{SM}) of Eq. \eqref{normal_ME}. In the stationary phase all magnetization components rapidly approach a constant value. Moreover, in spite of representing a single realization, they hardly display any fluctuations. The reason is that the stationary state is almost pure and an eigenstate of the jump operator, i.e., it satisfies to a very good approximation the relation 
\begin{equation}\label{state_below_threshold}
\hat{\rho}_\mathrm{ss}\approx|\beta\rangle\langle\beta|, \quad \hat{J}_- |\beta\rangle\approx-i\beta|  \beta\rangle 
\end{equation}
with $\beta=\omega N/(2\kappa)$. In fact, when approaching the thermodynamic limit the system features such pure stationary state in the entire interval $0\leq\omega/\kappa\leq1$ (see \cite{SM}). 

The trajectories shown in the top panel of Fig.~\ref{fig_regimes}(b) are not directly observable in experiment, as they entail the calculation of quantum expectation values $\langle\hat{m}_\alpha\rangle_\text{QJ}$ with the instantaneous wave function. By the nature of a quantum measurement such quantity cannot be obtained in a single shot. What is instead experimentally accessible --- e.g. in the experiment presented in  Ref. \cite{Ferioli2022} --- is the time record of emitted photons, which is shown in the middle panel of Fig.~\ref{fig_regimes}(b). Note, that this record corresponds to the same trajectory shown in the top row, which is also the case for panels (c) and (d). The rapid approach to stationarity is also visible here albeit fluctuations around the average value of the photon count (dashed black line) are clearly visible. Instead of counting photons one can monitor the so-called $x$-quadrature of the emitted light using a homodyne detection scheme \cite{Wiseman2009} (see \cite{SM}), which measures the homodyne current $I_\mathrm{x}(t)=\sqrt{2\kappa N}\langle \hat{m}_x(t)\rangle_\mathrm{H}+dW(t)/dt$. This observable is proportional to the instantaneous $x$-magnetization in the homodyne unravelling, indicated by the subscript $\text{H}$, plus the derivative of a random (Wiener) process $W(t)$ \cite{Wiseman2009,SM}. Removing this noise, via the application of a sliding average over the time windows $\kappa\Delta t=0.5$, yields the trajectories shown in the bottom row of  Fig. \ref{fig_regimes}(b-d). 

Let us now turn to Fig.~\ref{fig_regimes}(c), which shows the quantum trajectories at the critical point $\omega/\kappa=1$. Here the dynamics is strikingly different. All observables display a behavior in which periods of time, in which the quantum trajectories feature small fluctuations and drifts, are interspersed with sudden large fluctuations (marked with ${\bold\ast}$). This peculiar behavior, signatures of which are also found in the photon count and --- more clearly --- in the homodyne current, will be further analyzed below. We will find that the mean time between two consecutive large fluctuation events follows a power-law which is characteristic for critical phenomena. 

In the time-crystal phase the quantum trajectories of the magnetization exhibit non-decaying oscillatory behavior, as can be seen in  Fig.~\ref{fig_regimes}(d). Noise due to quantum jumps leads, however, to the relative dephasing of different trajectories, so that the average state displays exponentially damped oscillations \cite{Iemini2018}. This dephasing is observable in the oscillations of the magnetization when comparing them with the shadowed background that alternates in time intervals that are half the mean-field period. Note, that the larger the number of particles, the weaker the dephasing effect until persistent oscillations are achieved in the mean-field limit \cite{Iemini2018,Buonaiuto2021,Carollo2022}. Interestingly, also the photon count signal displays noisy but pronounced oscillations and thus allows for an \textit{in situ} detection of the time-crystal phase. In fact, the Fourier transform of the counting signal displays a peak at the mean-field frequency \cite{SM}. This peak becomes better resolved increasing system size as time-crystal oscillations become less noisy \cite{SM}. The time-crystal oscillations are also resolved in the (time-averaged) homodyne current shown in the bottom panel of Fig.~\ref{fig_regimes}(d).

\begin{figure}[t]
\includegraphics[width=1\linewidth]{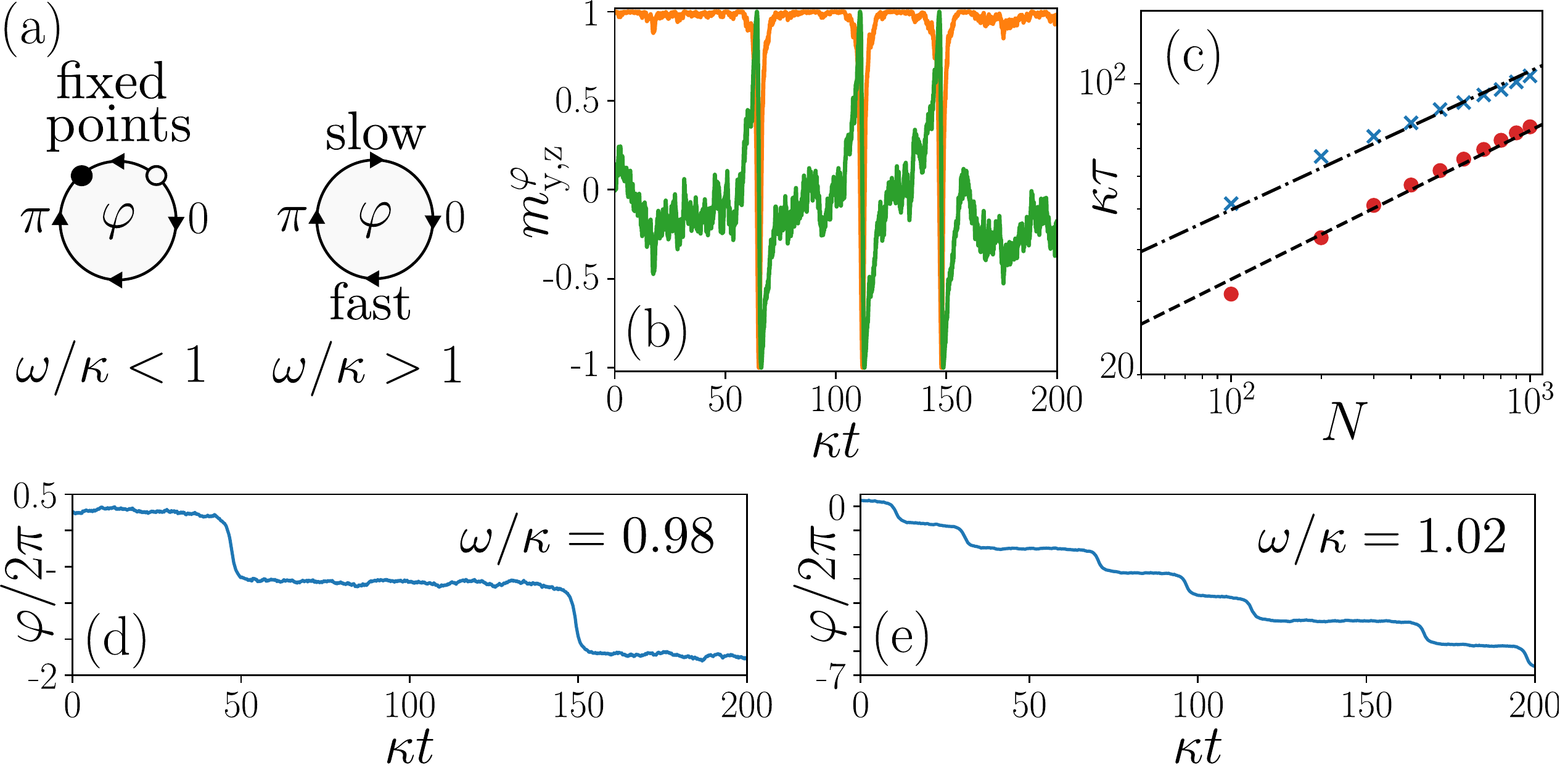}\\
\caption{\textbf{Dynamics near the critical point.} (a)  Flow of the phase equation below and above the critical point, where  $\bullet$ ($\circ$) denotes a (un-)stable fixed point (empty full bullet point). (b) Single realization of Eq. (\ref{toy_model}) with initial condition $\varphi=\pi/2$, $N=200$ and $\omega/\kappa=1$. (c) Average time between large fluctuations for $\omega/\kappa=1$ and varying $N$. Red circles correspond to the classical model fitted by  $\kappa\tau\propto N^{0.36}$ (black dashed line). Blue crosses correspond to the quantum jump process fitted by $\kappa\tau\propto N^{0.33}$ (dashed-dotted black line).  (d)-(e) Classical phase dynamics for two different frequencies.}\label{fig_class_model}
\end{figure}

\noindent {\it Dynamics near the critical point.--} In Fig.~\ref{fig_regimes}(c) we have shown that the dynamics of the quantum trajectories at the critical point is rather peculiar. To understand this behavior we return to the mean-field equations for the magnetization and augment the corresponding equations with Gaussian noise. We consider the case in which $m_x(0)=0$, which implies that the conserved quantity $\mathcal{M}$ takes the value zero. Defining the phase variable $\varphi$ via $m^\varphi_\mathrm{y}=\sin \varphi$ and $m^\varphi_\mathrm{z}=\cos \varphi$ leads to the following equation of motion:
\begin{equation}\label{toy_model}
\dot{\varphi}(t)=-\omega+\kappa\sin\varphi(t)+ \xi(t). 
\end{equation}
Here the Gaussian noise $\xi(t)$ is introduced to model the effect of finite-$N$ fluctuations and is characterized by the average value $\langle \xi(t)\rangle=0$ and the correlation function $\langle \xi(t)\xi(t')\rangle=\frac{2}{N}\delta(t-t')$. The pre-factor of the latter is chosen such that the noise vanishes in the thermodynamic limit.

The phase portrait of the deterministic part of this equation is shown in Fig.~\ref{fig_class_model}(a). Below the critical point there are one stable and one unstable fixed point which approach each other when $\omega$ increases. They coalesce and annihilate when the critical point, $\omega/\kappa=1$, is reached, for which the solution of the deterministic equation becomes a limit cycle. This time-periodic solution is non-harmonic, as the velocity of the phase-evolution depends on $\varphi$. In particular, around $\varphi\approx \pi/2$ this velocity decreases and becomes zero when the critical point is approached from above. This means in turn, that the period of the oscillation diverges \cite{Strogatz2018}. The proximity of the two fixed points slightly below criticality and the slow-down of the dynamics near $\varphi\approx \pi/2$ explain the large fluctuations observed in Fig.~\ref{fig_class_model}(b): slightly below the critical point, the noise occasionally allows the system to reach the unstable fixed point. After that $\varphi$ may immediately complete a full cycle in clockwise direction [Fig.~\ref{fig_class_model}(d)], manifesting as large fluctuation $m^\varphi_\mathrm{y,z}$ 
On the other hand, slightly above the critical point, the noise can  allow $\varphi$ to escape the region with small velocity, so that it undergoes afterwards a fast revolution, which also manifests as a large fluctuation $m^\varphi_\mathrm{y,z}$ [see Fig.~\ref{fig_class_model}(e)]. The average time $\tau$ between such large fluctuations as a function of $N$ is shown in Fig.~\ref{fig_class_model}(c) (red circles). As it can be seen, it obeys a power-law, $\tau\propto N^{0.36}$, which is characteristic for critical phenomena. The same power-law (within a reasonable margin of error) is observed for the average time between large fluctuations in the time-crystal trajectories.

\begin{figure}[t]
\includegraphics[width=1\linewidth]{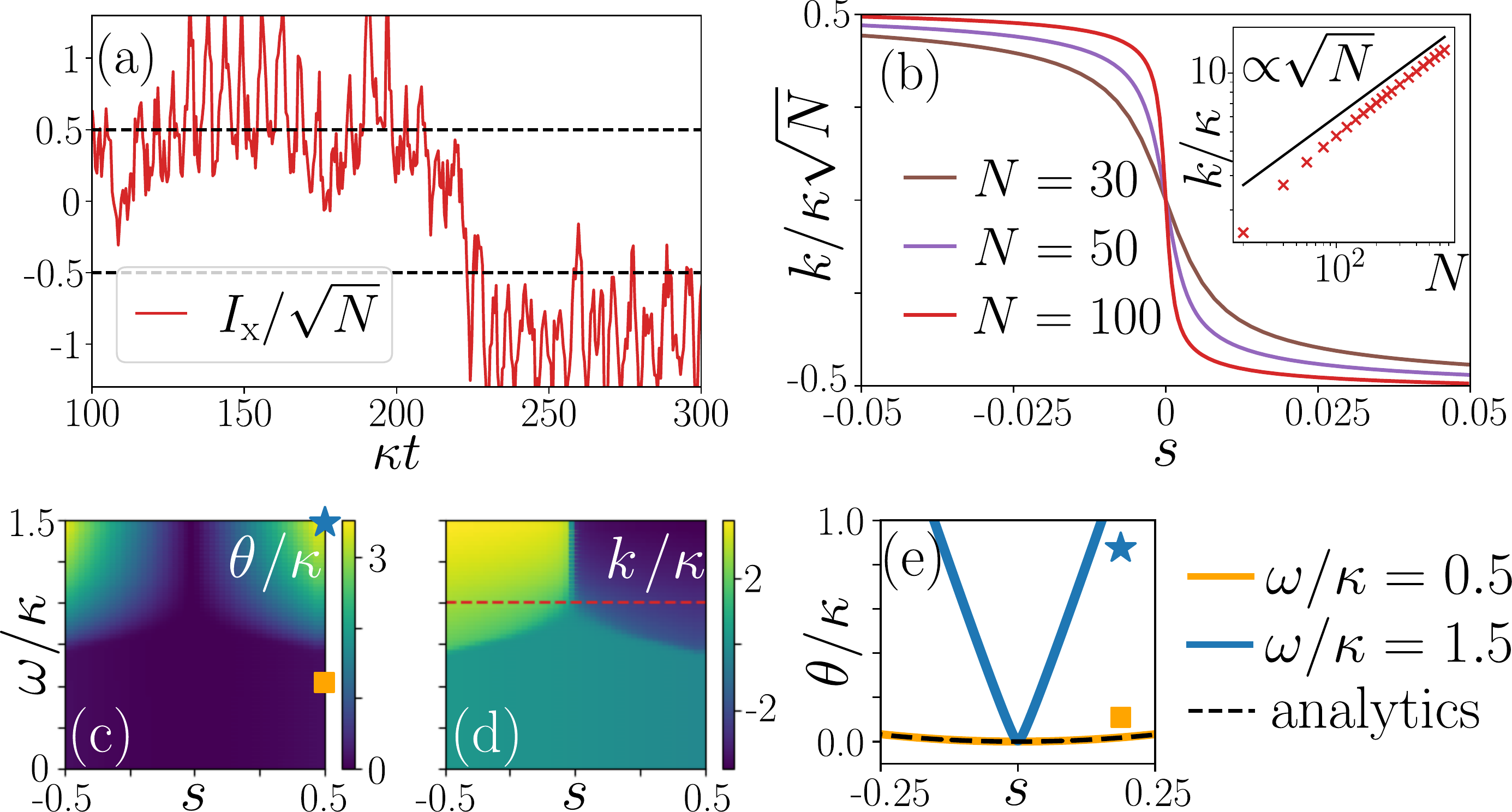}\\
\caption{\textbf{Dynamical order parameter.} (a)  Rescaled homodyne current for $\omega/\kappa=1.5$, $N=100$ and initial condition a coherent state with $\theta=\pi/2$ and $\phi=\pi/2$. (b) Rescaled activity $k(s)$ for various system sizes and $\omega/\kappa=1.5$. Inset: scaling of the activity at $s=-0.025$ and $\omega/\kappa=1.5$  with system size. (c,d) Colormaps for the scaled cumulant generating function $\theta(s)$ and the activity $k(s)$ for the homodyne current. (e) Scaled cumulant generating function $\theta(s)$ for two different values of $\omega/k$, one above and one below the critical point.} \label{fig_order_parameters}
\end{figure}
\noindent {\it Dynamical coexistence of oscillation patterns.--} In the following, we will show that in the time-crystal phase, the dynamics is comprised of different oscillatory patterns which dynamically coexist. This means that the quantum state features a given periodic solution for a long time-window but then eventually jumps at a random time into another one. This persistent jumping is related to a first-order dynamical phase transition \cite{Garrahan2010,Ates2012,Carollo2018}. To understand this, one has to keep in mind that, for the system in the thermodynamic limit, the breaking of the time-translation symmetry in the time-crystal regime results in the stabilization of several different oscillatory solutions, which are ``approached" according to the specific initial condition for the system state. However, when considering a finite system (subject to quantum fluctuations), the (long-time) state of the system is actually time-translation symmetric indicating that the latter must  consist of the ``sum" of the different oscillatory patterns \cite{Carmichael1980}. This \emph{statistical mixture of oscillations} is realized by the coexistence of all the possible oscillatory patterns in single homodyne trajectories. 

The above phenomenology is clearly displayed by single dynamical realizations of the homodyne current --- in the time-crystal regime --- when looking at sufficiently large time-scales as shown in Fig.~\ref{fig_order_parameters}(a). Here, the homodyne current shows oscillations which switch from a pattern with positive average value to a pattern with a negative one. On average the overall current must average to zero due to the expected value $\langle \hat{m}_{\mathrm{x}}\rangle_\mathrm{ss}=0$. One way to describe the emergence of this coexistence regime is achieved by using a ``thermodynamic formalism" for the full statistical characterization of the probability $p_K(t)$ of observing a given value of the time integrated homodyne current $K_t=\int_0^t dt' I_{\mathrm{x}}(t')$. Such characterization can be obtained via the moment generating function $Z(s)=\sum_{K=0}^\infty e^{-sK} p_K(t)$. For large times $t$, this function  behaves as $Z(s)\approx e^{t\theta(s)}$, where $\theta(s)$ is the so-called scaled cumulant generating function for $K_t$. This function can be computed as the largest real eigenvalue of the tilted generator \cite{Garrahan2010,Hickey2012,Carollo2021}
\begin{equation}\label{quarature_ME}
\mathcal{L}_s\hat{\rho}=\mathcal{L}\hat{\rho}-s\sqrt{\frac{2\kappa}{N}}\big(\hat{J}_-\hat{\rho}+\hat{\rho}\hat{J}_+  \big)+\frac{s^2}{2}\hat{\rho}\, .
\end{equation}
For instance, the average value of $K_t$ is given by the first derivative $k(s)=-\partial_s \theta(s)$,  calculated for $s=0$. When considered as a function of $s$, the scaled cumulant generating function $\theta(s)$ plays the role of a dynamical free energy, while the activity $k(s)$ can be regarded as an order-parameter which can signal emergent dynamical behavior in quantum trajectories \cite{Garrahan2010,Ates2012,Genway2012,Lesanovsky2013}. This is for instance evident in Fig.~\ref{fig_order_parameters}(b), where we plot the function $k(s)$ rescaled by $1/\sqrt{N}$. The activity $k(0)$ provides the average value of the time-integrated current in the typical trajectories, which is zero since there appear oscillatory patterns which are symmetric around $\langle \hat{m}_{\mathrm{x}}\rangle_{\mathrm{ss}}=0$. Around $s=0$, we see that the activity displays a rapid crossover from a phase with positive homodyne current to a phase with negative homodyne current. Such a crossover tends to approach a discontinuous transition when $N$ is increased. As for equilibrium phase transitions, this (almost) discontinuous behavior signals that in the typical trajectories, i.e., at $s=0$, there emerges coexistence of the two different phases, which here takes place in single stochastic realizations of the dynamics. It is possible to look at the dynamics for $s\neq0$ (see \cite{SM}). Here, the coexistence behavior is resolved by the parameter $s$, which plays the role of a biasing field, and single trajectories show a unique stable oscillatory pattern for sufficiently large  $|s|$. 

Such a critical behavior of the functions $\theta(s)$ and $k(s)$ disappears below the critical point, as shown in Fig.~\ref{fig_order_parameters}(c-d). In this case, the system approaches a unique stable stationary state characterized by vanishing instantaneous value of the average homodyne current and by small fluctuations. In particular, exploiting the approximate relation in Eq.~\eqref{state_below_threshold}, we can obtain the analytical expression 
$\theta(s)\approx s^2/2$
that we display in Fig.~\ref{fig_order_parameters}(e) together with the numerical result. We notice that this expression for $\theta(s)$ comes exclusively from the last term in the tilted generator in Eq.~\eqref{quarature_ME}, suggesting that the output homodyne current is essentially a pure white noise. This is in contrast with what happens above the critical point [cf.~Fig.~\ref{fig_order_parameters}(e)], where $\theta(s)$ becomes singular. 

{\it Summary and conclusions. --} Motivated by recent experimental progress, we analyzed signatures of a time-crystal phase transition in experimentally accessible quantum jump and homodyne trajectories. We illustrated how stationary and oscillatory phases manifest in these trajectories and unveiled peculiarities of the dynamics at the critical point, such as the occurrence of large fluctuations. Here we could show that the average time between two fluctuation events displays a power-law scaling with system size. We moreover showed that the time-integrated photo-current can serve as a dynamical order parameter for the time-crystal phase-transition. From this perspective the time-crystal phase can be regarded as a phase in which many oscillatory patters coexist. This dynamical phase coexistence manifests in strongly intermittent behavior of the photo-current. It would be interesting to understand, whether the dynamical transition may be useful as a resource for quantum enhanced metrology, as discussed in Ref. \cite{Macieszczak2016}.

{\it Acknowledgements. --} We are grateful for financing from the Baden-W\"urttemberg Stiftung through Project No.~BWST\_ISF2019-23. We also acknowledge funding from the Deutsche Forschungsgemeinschaft (DFG, German Research Foundation) under Project No. 435696605 and through the Research Unit FOR 5413/1, Grant No. 465199066. This project has also received funding from the European Union’s Horizon Europe research and innovation program under Grant Agreement No. 101046968 (BRISQ). F.C.~is indebted to the Baden-W\"urttemberg Stiftung for the financial support by the Eliteprogramme for Postdocs.

\bibliographystyle{apsrev4-2}
\bibliography{references.bib}

\setcounter{equation}{0}
\setcounter{figure}{0}
\setcounter{table}{0}
\makeatletter
\renewcommand{\theequation}{S\arabic{equation}}
\renewcommand{\thefigure}{S\arabic{figure}}

\makeatletter
\renewcommand{\theequation}{S\arabic{equation}}
\renewcommand{\thefigure}{S\arabic{figure}}

\onecolumngrid
\newpage

\setcounter{page}{1}

\begin{center}
{\Large SUPPLEMENTAL MATERIAL}
\end{center}
\begin{center}
\vspace{0.8cm}
{\Large Quantum trajectories of dissipative time-crystals}
\end{center}
\begin{center}
Albert Cabot$^{1}$, Leah Sophie Muhle$^{1}$, Federico Carollo$^{1}$, Igor Lesanovsky$^{1,2}$
\end{center}
\begin{center}
$^1${\it Institut f\"ur Theoretische Physik, Universit\"at T\"ubingen,}\\
{\it Auf der Morgenstelle 14, 72076 T\"ubingen, Germany}\\
$^2${\it School of Physics and Astronomy and Centre for the Mathematics and Theoretical Physics of Quantum Non-Equilibrium Systems, The University of Nottingham, Nottingham, NG7 2RD, United Kingdom}
\end{center}

\section{Monitoring processes}\label{SM}

In this section we discuss in more detail the two  unravellings of the master equation (\ref{normal_ME}) in quantum trajectories considered in this work: quantum jumps and homodyne detection. In both cases, we assume  the (ideal) conditions in which we are able to perfectly monitor the emitted light by the collective spin system in these two different ways.

{\it Quantum jump trajectories. --} By using a photon counting detection scheme, we have  access to the full record of times in which a photon emitted by the atomic system is detected. Then, we can unravel the master equation (\ref{normal_ME}) in stochastic trajectories for the state of the system {\it conditioned} on each detection time record. These are the so called quantum jump trajectories \cite{Wiseman2009,Gardiner2004}, and they can be modeled as follows. Starting from a pure state initial condition $|\Psi(t_0)\rangle$, the conditioned state evolves according to a non-Hermitian effective Hamiltonian:
\begin{equation}\label{non-Hermitian_evolution}
\frac{d}{dt}|\tilde{\Psi}_{\mathrm{QJ}}(t)\rangle=-i\big[\omega\hat{J}_{\mathrm{x}}-i\frac{\kappa}{N}\hat{J}_+\hat{J}_-  \big] |\tilde{\Psi}_{\mathrm{QJ}}(t)\rangle,
\end{equation}
where the tilde is used to denote an unnormalized state. If a photon is detected at time $t_j$,  the conditioned state of the system experiences a sudden jump:
\begin{equation}\label{quantum_jump}
|\Psi_{\mathrm{QJ}}(t_\mathrm{j}+dt)\rangle=\frac{\hat{J}_-|\tilde{\Psi}_{\mathrm{QJ}}(t_\mathrm{j})\rangle}{\sqrt{\langle \hat{J}_+\hat{J}_-\rangle_\mathrm{QJ}}},
\end{equation}
which gives the initial condition for the following period of non-Hermitian time evolution until the next detection occurs at time $t_{j+1}$. Here we have used the notation: $\langle \dots\rangle_\mathrm{QJ}= \langle {\Psi}_{\mathrm{QJ}}(t_{\mathrm{j}})|\dots|{\Psi}_{\mathrm{QJ}}(t_\mathrm{j})\rangle$. On average over realizations this time evolution is equivalent to the master equation (\ref{normal_ME}). However, individual quantum trajectories simulate what would be observed in individual experimental runs under such ideal conditions.

{\it Homodyne trajectories. --} This is a different possible unravelling of the master equation (\ref{normal_ME}), which models idealized homodyne detection of the emitted light \cite{Wiseman2009}. Moreover, we assume that we are performing this kind of monitoring in such a way that corresponds to the x-quadrature of the spin system. Then, our dynamics is described by the following stochastic non-linear equation for the conditioned state \cite{Wiseman2009}:
\begin{equation}\label{homodyne_equation}
\begin{split}
d|\Psi_\mathrm{H}(t)\rangle=\bigg\{-i\omega\hat{J}_\mathrm{x}dt-\frac{\kappa}{N}\big[\hat{J}_+\hat{J}_- 
-\langle\hat{x}(t)\rangle_\mathrm{H}\hat{J}_-
+\langle\hat{x}(t)/2\rangle_\mathrm{H}^2\big]dt
+\sqrt{\frac{2\kappa}{N}}[\hat{J}_- -\langle\hat{x}(t)/2\rangle_\mathrm{H}]dW(t)\bigg\}|\Psi_\mathrm{H}(t)\rangle,
\end{split}
\end{equation}
where $\hat{x}=(\hat{J}_++\hat{J}_-)$, and $dW(t)$ is a Wiener increment. Moroevorer, we have used the notation  $\langle \dots \rangle_\mathrm{H}=\langle\Psi_\mathrm{H}(t)|\dots|\Psi_\mathrm{H}(t)\rangle$. In this kind of monitoring the output measurable quantity is the homodyne current  or photo-current, given by:
\begin{equation}\label{homodyne_current}
I_\mathrm{x}(t) =\sqrt{\frac{2\kappa}{N}}\langle\hat{x}(t)\rangle_\mathrm{H}+\frac{dW(t)}{dt}.  
\end{equation}

\subsection{Mean-field solutions}

In this section we provide some further details on the mean-field dynamics of the system that emerges in the thermodynamic limit. As commented in the main text, the mean-field description is characterized by two conserved quantities: the conservation of the norm $m_\mathrm{x}^2+m_\mathrm{y}^2+m_\mathrm{z}^2=1$ which originates from the conservation of  $j^2$ in the full quantum model; and the conservation of $\mathcal{M}$, which emerges as the thermodynamic limit is approached. Below the critical point, there is a stable fixed point attractor given by
\cite{Carmichael1980,Hannukainen2018}:
\begin{equation}\label{fixed_point}
m_\mathrm{x}=0, \quad m_\mathrm{y}=\frac{\omega}{\kappa}, \quad m_{z}=-\sqrt{1-\frac{\omega^2}{\kappa^2}}.
\end{equation}
Above the critical point,  $\omega/\kappa>1$, the system displays non-decaying oscillations. As commented in the main text, the conserved quantity $\mathcal{M}$ leads to a continuous family of non-isolated closed orbits covering all Bloch sphere \cite{Drummond1978,Carmichael1980,Iemini2018}. Therefore, the oscillatory regime of this driven-dissipative system is rather different from stable limit-cycles (or isolated closed orbits) found in typical non-equilibrium scenarios \cite{Strogatz2018}, and it instead resembles Hamiltonian dynamics. As noticed in Ref. \cite{Iemini2018}, the mean-field equations define a {\it reversible} dynamical system. In this peculiar scenario \cite{Roberts1992}, non-Hamiltonian dynamical systems can display both the typical attractors of dissipative non-linear systems as well as continuous families of closed periodic orbits. Analytical expressions for these periodic trajectories were obtained in Refs. \cite{Drummond1978,Carmichael1980,Carollo2022}. Focusing for simplicity on $\mathcal{M}=0$,  this solution reads \cite{Carmichael1980,Carollo2022}:
\begin{equation}\label{time_crystal_MF}
\begin{split}
m_\mathrm{x}(t)=0\quad
m_\mathrm{y}(t)=\frac{\omega}{\kappa}+\frac{(\omega/\kappa)^2-1}{\cos(\Omega t -\phi_0)-\omega/\kappa},\quad
m_\mathrm{z}(t)=\frac{\Omega \sin(\Omega t-\phi_0)}{\kappa\cos(\Omega t-\phi_0)-\omega},
\end{split}
\end{equation}
with $\Omega=\sqrt{\omega^2-\kappa^2}$, $\sin\phi_0=[\Omega m^0_\mathrm{z}/(\omega-\kappa m^0_\mathrm{y})]$, $\cos\phi_0=[1-(\omega/\kappa)m^0_\mathrm{y}]/(\omega/\kappa- m^0_\mathrm{y})$, and $m^0_{\mathrm{y,z}}$ are the initial conditions. We notice that for all values of $\mathcal{M}$ the fundamental oscillation frequency is $\Omega$ \cite{Drummond1978,Carmichael1980}, which becomes zero at the transition point, resembling an infinite-period bifurcation \cite{Strogatz2018}.

\section{Stationary state below the critical point}

\begin{figure}[t]
\includegraphics[width=0.6\linewidth]{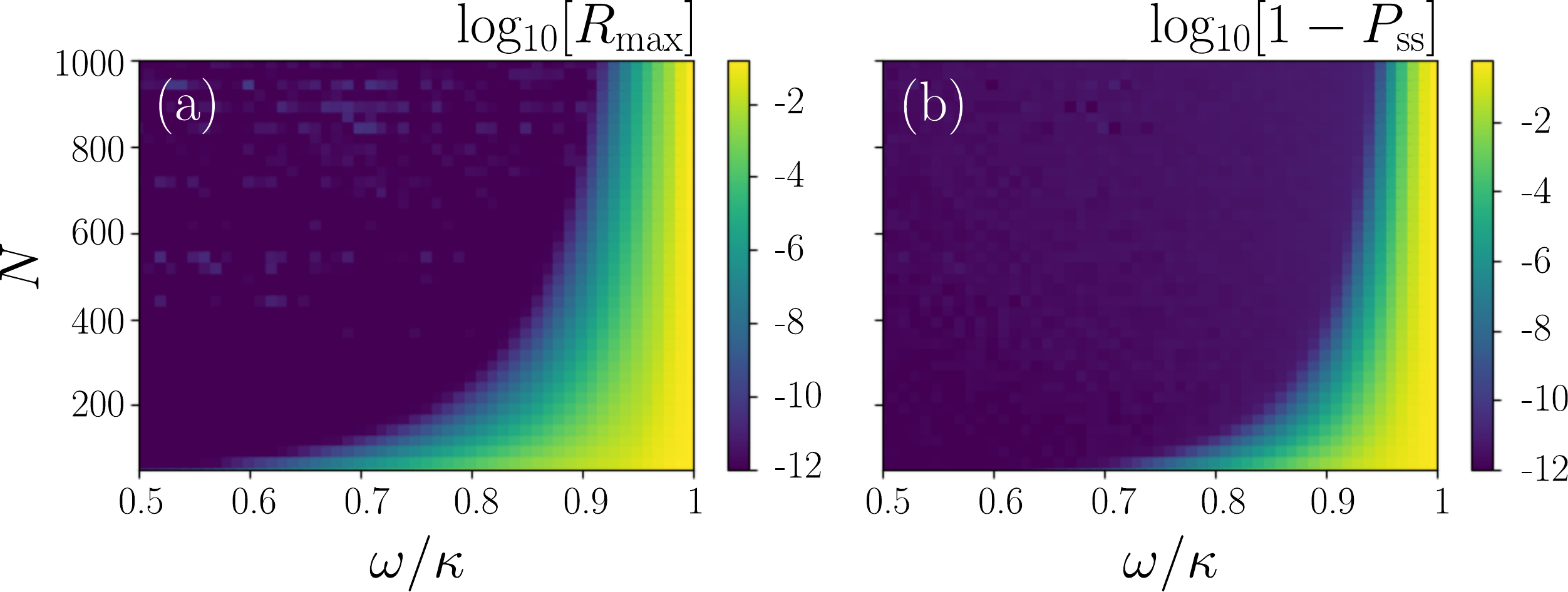}\\
\caption{\textbf{Stationary state below the critical point.} (a) In color: logarithm of $R_{\mathrm{max}}$ defined as the maximum in absolute value matrix element of $\hat{J}_-\hat{\rho}_\mathrm{ss}-\langle \hat{J}_-\rangle_{\mathrm{ss}}\hat{\rho}_\mathrm{ss}$. (b) Logarithm of 1 minus the purity of the stationary state.}\label{fig_state}
\end{figure}

In this section we analyze the accuracy of Eq. (\ref{state_below_threshold}). We then consider two figures of merit for the actual stationary state of the master equation $\hat{\rho}_\mathrm{ss}$. The first one is its purity, as its approximation by $|\beta\rangle\langle\beta|$ assumes the stationary state to be pure. The second one is more direct, and it checks whether the stationary state is invariant under the action of $\hat{J}_-$. In particular we define the operator:
\begin{equation}
\hat{R}=\hat{J}_-\hat{\rho}_\mathrm{ss}-\langle \hat{J}_-\rangle_{\mathrm{ss}}\hat{\rho}_\mathrm{ss}.
\end{equation}
If Eq. (\ref{state_below_threshold}) was exact, then all matrix elements of $\hat{R}$ would be zero. Thus, a way to analyze the accuracy of this approximation is computing the maximum (in absolute value) matrix element of $\hat{R}$ which we call $R_\mathrm{max}$:
\begin{equation}
R_\mathrm{max}=\max_{\forall i,j}\big[|\big(\hat{J}_-\hat{\rho}_\mathrm{ss}-\langle \hat{J}_-\rangle_{\mathrm{ss}}\hat{\rho}_\mathrm{ss}\big)_{ij}|\big]    
\end{equation}
These figures of merit are plotted in Fig. \ref{fig_state} varying $\omega/\kappa$ and $N$. In panel (a) we plot $R_\mathrm{max}$ while in (b) we plot 1 minus the purity of the stationary state. The dark blue regions correspond to the parameter regions in which the approximation (\ref{state_below_threshold}) is very accurate. We see that these regions grow towards the mean-field transition point, $\omega/\kappa=1$, as $N$ is increased. Indeed,  the analysis of Fig. \ref{fig_state} provides a means to the distinguish more clearly between the dynamics well-below and near the critical point for finite $N$; as long as one lies within the dark blue regions the system  displays the kind of dynamics shown in Fig. \ref{fig_regimes} (b), while out of these regions the dynamics looks like Fig. \ref{fig_regimes} (c) in which quantum jumps affect significantly the state of the system since this no longer satisfies to a good approximation Eq. (\ref{state_below_threshold}).

\section{Doob transformed homodyne master equation}

\begin{figure}[t]
\includegraphics[width=0.9\linewidth]{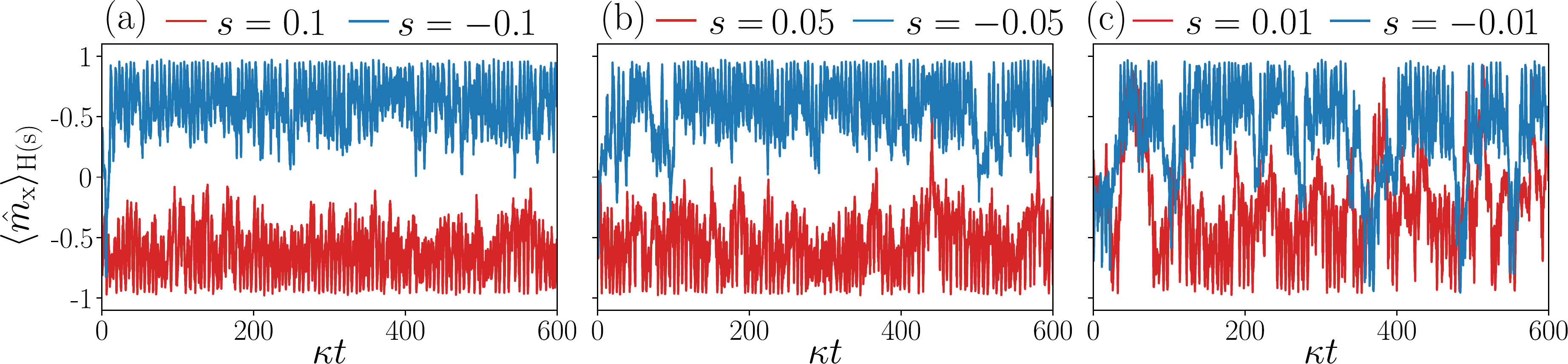}\\
\caption{\textbf{Homodyne trajectories for the Doob transformed master equation.} Magnetization x-components for the Doob transformed system and different values of the tilting field: (a) $s=\pm0.1$; (b) $s=\pm0.05$; (c)  $s=\pm0.01$. Parameters: $\omega/\kappa=1.5$, $N=30$ and initial condition $\theta=\pi/2$ and $\phi=\pi/2$.}\label{fig_MF_comparison}
\end{figure}

We can design a physical system for which its typical realizations in, e.g., homodyne trajectories correspond to rare events of our original system \cite{Carollo2018}. This is achieved by means of the Doob transform of the tilted master equation for the corresponding observation process \cite{Carollo2018,Carollo2021}. This recipe makes use of the leading eigenvalue of the tilted generator and the corresponding {\it left} eigenmatrix. Here we are interested in the Doob transform of the homodyne monitored system, as we want to further illustrate the physical meaning of the results for $s\neq0$. Therefore the starting point are the leading eigenvalue/eigenmatrices of the tilted operator for the homodyne process:
\begin{equation}
\mathcal{L}_s\hat{R}_0=\theta(s)\hat{R}_0,\quad \mathcal{L}^\dagger_s \hat{L}_0=\theta(s)\hat{L}_0,
\end{equation}
which satisfy:
\begin{equation}
\hat{R}_0^\dagger=\hat{R}_0, \quad \hat{L}_0^\dagger=\hat{L}_0,\quad \text{Tr}[\hat{R}_0]=1,\quad   \text{Tr}[\hat{L}_0\hat{R}_0]=1.  
\end{equation}
Then the time-evolution of the Doob transformed system is given by:
\begin{equation}
\frac{d}{dt}\hat{\rho}=\mathcal{L}_\mathrm{D}\hat{\rho}.    
\end{equation}
Note that this corresponds to a trace preserving completely positive time-evolution, which thus can be written in the standard Lindblad form. For the homodyne tilted master equation we can obtain the Doob transformed system as \cite{Garrahan2010,Carollo2018,Carollo2021}:
\begin{equation}
\mathcal{L}_\mathrm{D}[\hat{\rho}]=\hat{L}_0^{1/2}  \mathcal{L}_s\big[\hat{L}_0^{-1/2}\hat{\rho} \hat{L}_0^{-1/2}\big]\hat{L}_0^{1/2}  -\theta(s)\hat{\rho}
\end{equation}
where in order to avoid confusion, in this equation we have written within square brackets $[\cdot]$ those operators that are acted by the superoperators $\mathcal{L}_\mathrm{D}$ or $\mathcal{L}_s$. Defining the transformed jump operator:
\begin{equation}
\tilde{J}_-=\hat{L}_0^{1/2} \hat{J}_-\hat{L}_0^{-1/2},   
\end{equation}
the Doob transformed system reads:
\begin{equation}
\mathcal{L}_\mathrm{D}\hat{\rho}=-i[\hat{H}_\mathrm{D},\hat{\rho}]+\frac{\kappa}{N}\big(2\tilde{J}_-\hat{\rho}\tilde{J}_+-\tilde{J}_+\tilde{J}_-\hat{\rho}-\hat{\rho}\tilde{J}_+\tilde{J}_-\big)    
\end{equation}
with
\begin{equation}
\hat{H}_\mathrm{D}=\omega \tilde{J}_\mathrm{x}-s\sqrt{\frac{2\kappa}{N}}\tilde{J}_\mathrm{y},    
\end{equation}
where $\tilde{J}_\mathrm{x}=(\tilde{J}_++\tilde{J}_-)/2$ and $\tilde{J}_\mathrm{y}=i(\tilde{J}_--\tilde{J}_+)/2$. Notice how in the Hamiltonian term there a appears a new coherent driving term with strength proportional to the tilting field $s$ and in the (transformed) $y$ direction. This master equation can also be unravelled in an homodyne stochastic process assuming that we are able to monitor the (transformed) x-component, i.e. $\tilde{J}_\mathrm{x}$. These homodyne trajectories are described by:
\begin{equation}\label{homodyne_equation_tilted}
\begin{split}
d|\Psi_\mathrm{H(s)}(t)\rangle=\bigg\{-i\hat{H}_\mathrm{D}(s)dt-\frac{\kappa}{N}\big[\tilde{J}_+\tilde{J}_- 
-\langle\tilde{x}(t)\rangle_\mathrm{H(s)}\tilde{J}_-
+\langle\tilde{x}(t)/2\rangle_\mathrm{H(s)}^2\big]dt
+\sqrt{\frac{2\kappa}{N}}[\tilde{J}_- -\langle\tilde{x}(t)/2\rangle_\mathrm{H(s)}]dW(t)\bigg\}|\Psi_\mathrm{H(s)}(t)\rangle,
\end{split}
\end{equation}
which is analogous to the previous homodyne process but replacing the jump operators by the transformed ones, and adding the extra Hamiltonian term. Notice that we have equivalently defined: $\tilde{x}=(\tilde{J}_++\tilde{J}_-)$ and $\langle \dots \rangle_\mathrm{H(s)}=\langle\Psi_\mathrm{H(s)}(t)|\dots|\Psi_\mathrm{H(s)}(t)\rangle$.

In Fig. \ref{fig_MF_comparison} we exemplify trajectories for the x-component magnetization and different values of $s$ in the time-crystaline regime with $\omega/\kappa=1.5$. As we can see, increasing $s$ has the effect of reducing the sign intermittency of the x-magnetization: the larger in absolute value the tilting field is, the more time typically spends the system without changing sign of the x-component. This is consistent with the fact that typical trajectories for the Doob transformed master equation correspond to rare events in the un-transformed master equation. Moreover, in agreement with the results of Fig. \ref{fig_order_parameters} (b), when tilting away from $s=0$ in the negative (positive) direction the activity steeply acquires a positive (negative) value and thus the  x-magnetization  spends more time in the positive (negative) axis.  Here we recall that we observe both $\langle \hat{J}_\mathrm{x} \rangle_\mathrm{H(s)}$ and $\langle \tilde{J}_\mathrm{x} \rangle_\mathrm{H(s)}$ to display a similar behavior.

\section{Frequency analysis of the counting signal}

\begin{figure}[t]
\includegraphics[width=0.9\linewidth]{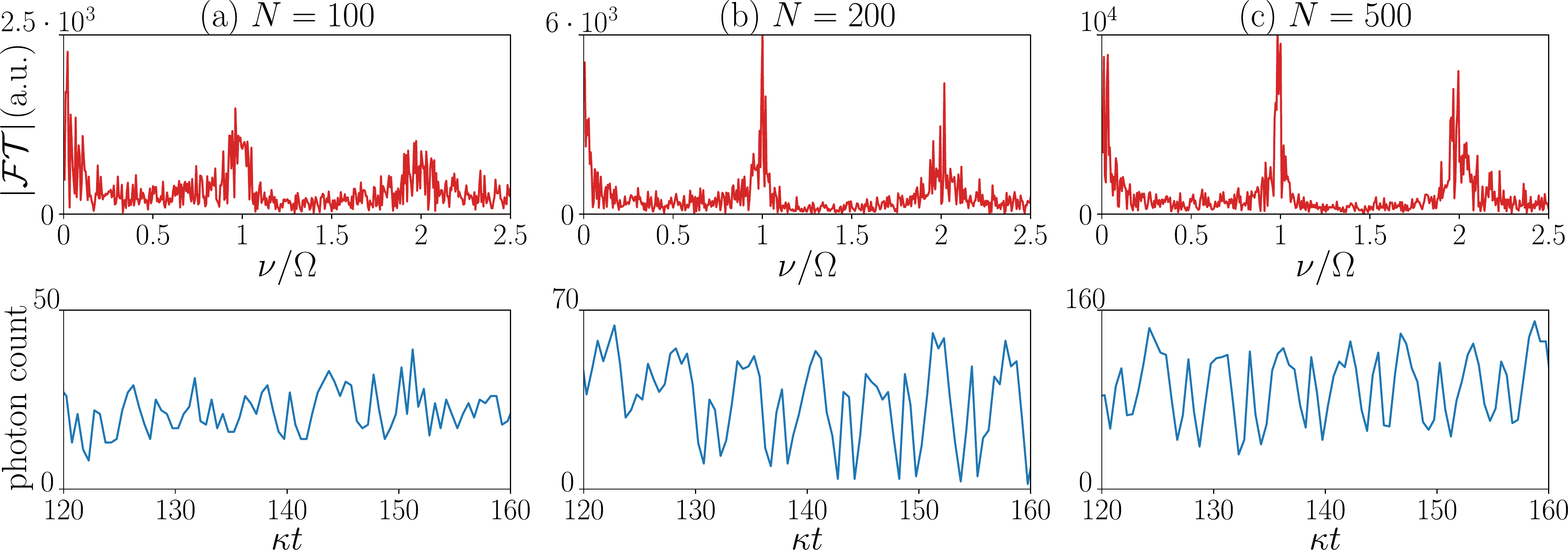}\\
\caption{\textbf{Frequency analysis of the counting signal.} Top rows: absolute value of the discrete Fourier transform of the counting signal. The frequency axis is given in units of the mean-field frequency $\Omega=\sqrt{\omega^2-\kappa^2}$. Bottom plots: small interval of the full counting signal used for the top panels (total length $\kappa t=1000$). Parameters: $\omega/\kappa=1.5$, and as initial condition a spin coherent state with $\theta=\pi/2$ and $\phi=\pi/2$. Counting signals binned in sliding time windows of $\kappa \Delta t=0.5$ as in the main text. (a) $N=100$, (b) $N=200$ and (c) $N=500$.  }\label{fig_FT}
\end{figure}

In this section we analyze in more detail the oscillatory signal in the counting record that we have reported in Fig. \ref{fig_regimes} (d) of the main text. In particular, in  Fig. \ref{fig_FT} we show the Fourier transform of the binned photon counting records for three different sizes (top row), while in the bottom row we show a small time interval of the corresponding counting signal. As we can see, the Fourier transform displays peaks at multiples of the fundamental mean-field frequency. Moreover, when increasing the system size these peaks become sharper and better resolved compared to the background noisy spectrum. This is in agreement with what we observe in the corresponding photon count record (lower row) in which cleaner oscillations are displayed for larger system sizes.

\section{Determining the time in between large fluctuations}

Here we provide the technical details on how we have obtained the data points of Fig. \ref{fig_class_model} (c) in the main text. In order to systematically analyse the spikes or large fluctuations of Figs. \ref{fig_regimes} (c) and  \ref{fig_class_model} (b) we need to define a threshold-value for which a fluctuation is considered large. For this task the most straightforward way is to consider the y-component as the large fluctuation for it corresponds to a simple dip from 1 to a small value. We then fix this threshold value at $m_\mathrm{y}=0.8$. Then we run long enough trajectories in order to get enough number of big fluctuations. In particular the values for the average times between the large fluctuations are obtained from data sets of at least $10^4$ large fluctuations.

\end{document}